\documentclass{elsart}

\usepackage{natbib}

\usepackage{graphics}
\usepackage{graphicx}
\usepackage{epsfig}

\usepackage{amssymb}

\begin{document}

\begin{frontmatter}



\title{MHD Interaction of Pulsar Wind Nebulae with SNRs and with the ISM}


\author{E. van der Swaluw}

\address{FOM-Institute for Plasma Physics Rijnhuizen, P.O. Box 1207, 3430 
BE Nieuwegein, The Netherlands 
}

\begin{abstract}
In the late 1960s the discovery of the Crab pulsar in its associated supernova 
remnant, launched a new field in supernova remnant research: the study of 
pulsar-driven or plerionic supernova remnants. In these type of remnants, 
the relativistic wind emitted by the pulsar, blows a pulsar wind nebula into 
the interior of its supernova remnant. Now, more then forty years after the 
discovery of the Crab pulsar, there are more then fifty plerionic supernova 
remnants known, due to the ever-increasing capacity of observational 
facilities. These observational studies reveal a Zoo of complex morphologies 
over a wide range of frequencies, indicating the significance of the 
interaction between a pulsar wind nebula with its surrounding supernova 
remnant. A pulsar which gained a kick velocity at birth, will ultimately break 
outside of its remnant, after which the pulsar wind nebula interacts directly 
with the interstellar medium. In general these pulsar wind nebulae are bounded 
by a bow shock, due to the supersonic motion of the pulsar.
There are a few examples known of these pulsar-powered bow shocks, a number 
which is slowly increasing.

I will review our current understanding of the different evolutionary stages of 
a pulsar wind nebula as it is interacting with its associated supernova remnant.
Therefore I will discuss both analytical and more recent numerical (M)HD models.
The four main stages of a pulsar wind nebula are: the supersonic expansion stage,
the reverse shock interaction stage, the subsonic expansion stage and ultimately
the stage when the head of the bubble is bounded by a bow shock, due to the
supersonic motion of the pulsar. Ultimately this pulsar wind nebula bow shock 
will break through its associated remnant, after which the pulsar-powered bow 
shock will interact directly with the interstellar medium. I will discuss 
recent numerical models from these type of pulsar wind nebulae and their 
morphology.
\end{abstract}


\end{frontmatter}

\section{Introduction}
\label{sect1}

After the explosion of a massive star, the outer layers of the star are
ejected into the circum-stellar or interstellar medium (ISM). This event gives
birth to a supernova remnant (SNR): an expanding bubble of million-degree gas.
A {\it plerionic} supernova remnant results from those explosion which also
yield a fast rotating neutron star, which blows a pulsar wind nebula (PWN) into
the interior of its remnant. Recent models by Heger et al. (2003) indicate
a mass range of the progenitor star between 10 to 25 solar masses, in order
to obtain both a supernova explosion and a neutron star.
The ejecta from the progenitor star are driven into the surrounding medium,
which at early stages yields a double-shock structure (McKee 1974; McKee and
Truelove, 1995): a forward shock propagating into the surrounding medium, 
and a reverse shock which forms {\it almost immediately}, shock-heating the 
freely expanding ejecta to X-ray emitting temperatures.Ultimately, as the 
forward shock has swept up a few times the ejecta mass, the reverse shock 
propagates towards the center of the remnant, where its energy is dissipated
into weak shock waves and sound waves (Cioffi et al. 1998).
In recent years the number of plerionic remnants has slowly been increasing
(see Kaspi et al. (2004) for a recent overview),
making this type of remnant a rather common type of SNR. The
presence of a PWN in the interior of a remnant obviously
increases the complexity of the internal dynamics of the remnant, although
the energy input from a pulsar wind over its lifetime is small, therefore its
evolution couples to the remnant's evolution. The PWN itself is
driven by a pulsar wind (Pacini and Salvati, 1973; Rees and Gunn, 1974), which is believed 
to be {\it highly} relativistic and magnetized (see e.g. Kennel and Coroniti, 1984; Gallant et al., 
2002). The wind itself consists of an electron-positron plasma, possibily together 
with an ion ingredient (Hoshino et al., 1992), and is ultimately terminated by
a strong MHD shock. At the site of the wind termination shock, particles are being
accelerated. These accelerated particles radiate part of their energy away as synchrotron 
radiation, while being advected away from the wind termination shock. These regions of
the PWN are observed over a wide range of frequencies as the actual plerion.

\section{Plerionic supernova remnants with a centered pulsar}
\label{sect2}

\subsection{Analytical models: supersonic and subsonic expansion}

The morphology of a PWN, as it is still interacting with the freely
expanding ejecta of its associated remnant, can be divided into three
regions separated by three interfaces (see left panel Figure 1): A/ the 
pulsar wind cavity, B/ the pulsar wind (synchrotron) bubble, and C/ the 
shell of swept-up ejecta material. Whereas the three interfaces are I/ 
the pulsar wind termination shock, II/ the contact discontinuity and 
III/ the PWN shock. The PWN
shock is propagating through the freely expanding ejecta (see right
panel Figure 1) and is being
accelerated ($R_{\rm pwn}\propto t^{6/5}$ for a uniform density of the ejecta) 
as long as the pulsar wind 
luminosity is constant (Reynolds \& Chevalier 1984, Chevalier \& Fransson
1992, van der Swaluw et al. 2001). 
This stage of the PWN evolution has been called {\it the supersonic
expansion stage} in literature and its timescale is determined by
{\it when} the reverse shock of the SNR collides with the PWN shock.
After the passage of the reverse shock, the expansion of the PWN
will be subsonic (Reynolds \& Chevalier 1984).

McKee and Truelove (1995) give analytical approximations for the 
trajectories of the forward shock and the reverse shock for the case
in which both the ambient medium and the ejecta have a uniform density.
The reverse shock reaches the center of the remnant at a timescale of
\begin{equation}
t_{\rm col}\;\simeq\; 1\; 045 E^{-1/2}_{51}
\left({M_{\rm ej}\over M_\odot}\right)^{5/6}n_0^{-1/3}\;\;{\rm years}\; ,
\end{equation}
here $E_{51}$ is the total mechanical energy of the SNR in units of $10^{51}$ 
erg, $n_0$ is the ambient hydrogen number density assuming an interstellar 
composition of 10 H : 1 He, and $M_{\rm ej}$ is the ejecta mass. The above
equation yields an upper limit for the timescale of the supersonic expansion
stage of the PWN. The associated maximum radius of the PWN can be calculated
using equation 12 from van der Swaluw et al. (2001), which yields:
\begin{equation}
R_{\rm col}\;\simeq\; 3.69\; L_{38}^{1/5}\; M_{\rm ej}^{1/2}\; n_0^{-2/5}\;{\rm parsec}\; ,
\end{equation}
here $L_{38}$ is the luminosity of the (constant) pulsar wind in units of 
$10^{38}$ ergs/sec.

The above approximations yield rough estimates for the timescale and the
maximum size of the PWN in the supersonic expansion stage, when both the 
ejecta and the ambient medium have a uniform density. 
A more sophisticated analysis would have to include 1/ a non-uniform density 
of the ejecta (Truelove and McKee, 1999) 2/ a non-uniform ambient medium which
has been  adapted by the activity of the wind from the progenitor star (see e.g. 
Garcia-Segura et al., 1996) and 3/ the decay of the pulsar wind luminosity
(see e.g. Bucciantini et al. 2004). Chevalier (2004) has recently used these 
three aspects into a model to deduce the supernova type of several observed 
SNRs.

\begin{figure}
\includegraphics[width=140mm]{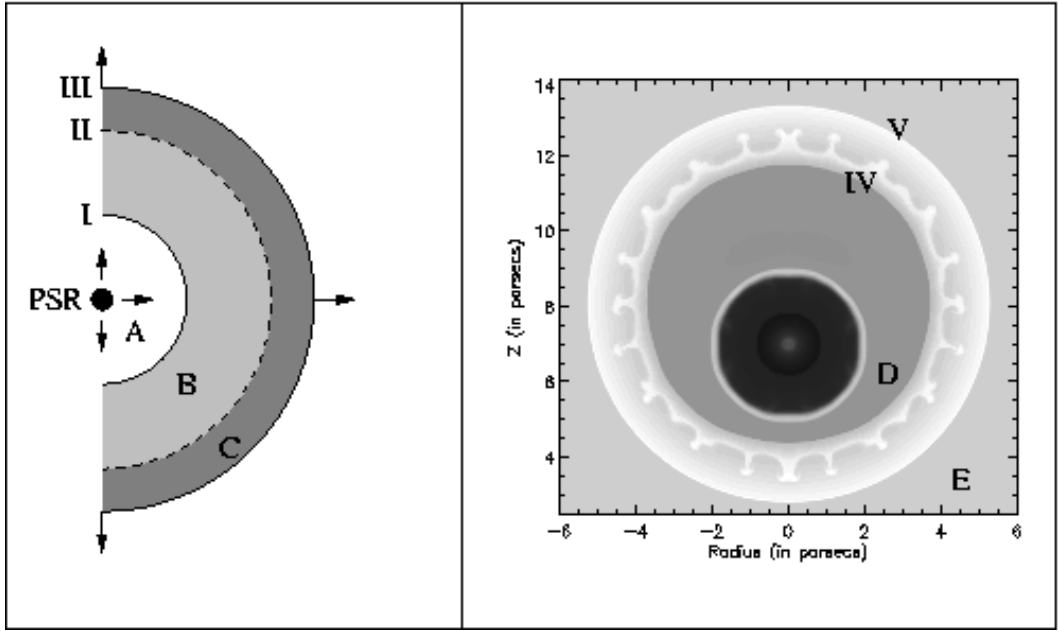}
\caption{The left panel shows a schematic representation of a PWN expanding
in the freely expanding ejecta. There are three interfaces: I the pulsar wind
termination shock; II the contact discontinuity dividing shocked pulsar wind 
material from shocked ejecta; III the PWN shock. Region A is the pulsar wind
cavity, region B is the pulsar wind bubble and region C is the shell
containing the swept-up ejecta. The right panel shows a logarithmic gray-scaling 
of the density distribution of a plerionic SNR from a hydrodynamical simulation 
(van der Swaluw et al. 2004). 
Apart from all the interfaces and 
regions from the left panel one can distinguish IV the reverse shock and V the forward
shock. Region D consists of freely expanding ejecta, whereas region E 
denotes the interstellar medium.}
\end{figure}

\subsection{Hydrodynamical simulations: the reverse shock interaction}

\begin{figure}
\includegraphics[width=140mm]{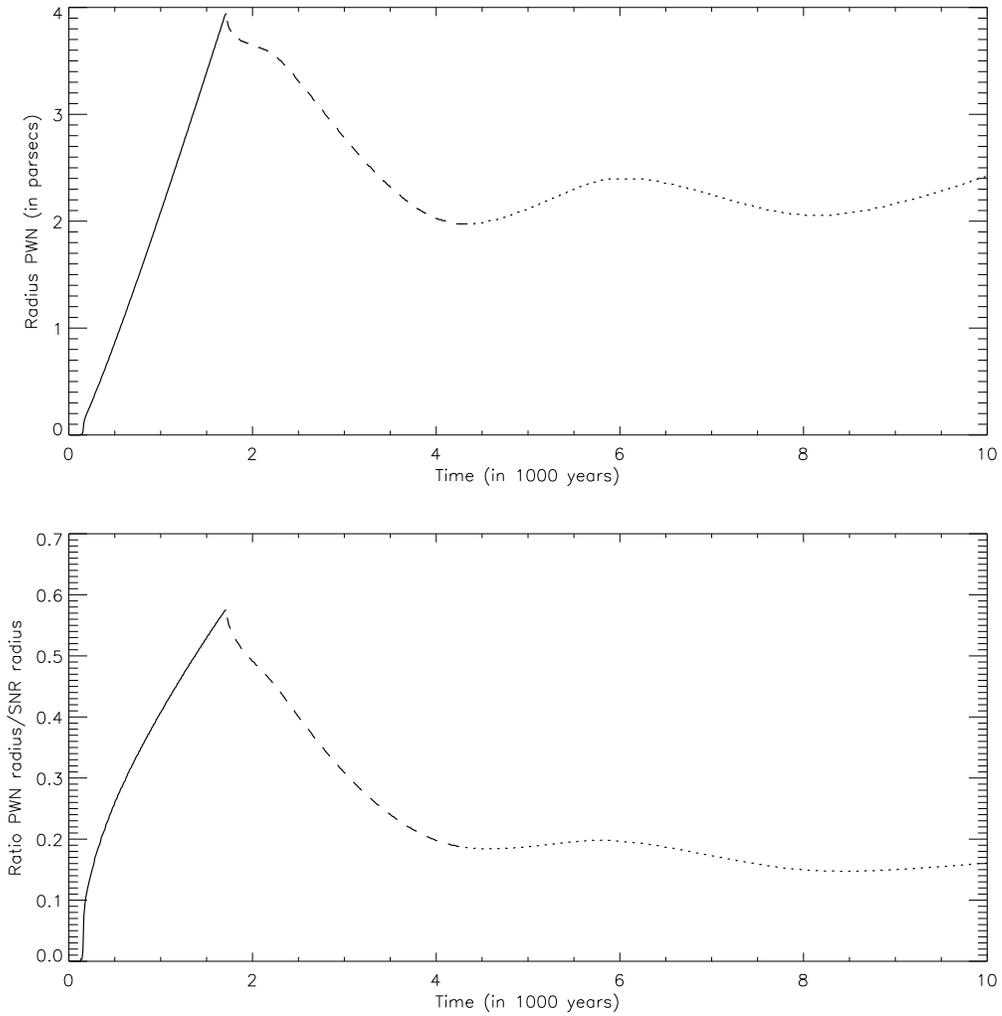}
\caption{The upper panel shows the radius of the PWN as a function of
time.  The lower panel shows the ratio of the PWN radius over the blastwave
radius from the SNR as a function of time. For both panels the solid line 
corresponds with the supersonic expansion stage, the dashed line corresponds 
with the crushing stage and the dotted line marks the stages in which the PWN 
relaxes towards the subsonic expansion stage. The simulation results are 
adapted from van der Swaluw et al. (2001).}
\end{figure}

In recent years numerical simulations have yielded a more sophisticated
model for plerionic supernova remnants with a centered pulsar 
($V_{\rm psr}=0$) and the associated reverse shock interaction. These 
simulations (e.g. van der Swaluw et al. 1998, 2001; Blondin et al. 2001; 
Bucciantini et al. 2003) confirm the division into an initial supersonic 
expansion stage and a subsequent subsonic expansion stage. However these simulations
all very clearly show that the transition between these two stages is not
direct, but via a non-steady expansion stage of the PWN: the PWN is  
crushed, shortly after the passage of the reverse shock, in order to gain
pressure balance with the surrounding SNR interior. The timescale of this
crushing event seems to be comparable to the lifetime of the supersonic 
expansion stage (see Figure 2), which implies that a significant 
amount of observed PWNe should be in this intermediate stage of their 
evolution. The characteristics of these type of crushed PWNe would be 
1/ small size of the PWN with respect 
to the SNR radius (see Figure 2), 2/ very bright emission due to the increase of the magnetic field strengths 
(caused by the decrease of the bubble's size), and 3/ a synchrotron cooling
break at relative low 
frequency due to the increasing magnetic field strength (Gallant et al., 2002). 
After the PWN has been crushed, the bubble oscillates between expansion and
contraction, due to the reverberations of the reverse shock interaction, 
after which the PWN relaxes towards a steady subsonic expansion. 

\subsection{Magneto-hydrodynamical simulations: elongated PWNe}

The first images of the Crab Nebula already revealed the elongation of
the pulsar wind bubble. It is very clear now that the Crab Nebula is not
unique but that this is a rather common aspect of supersonically
expanding PWNe. Begelman and Li (1998) developed a 
semi-analytical model to explain the elongation of the Crab Nebula as a 
result of the toroidal magnetic fields inside the PWN. The toroidal 
magnetic fields introduce a pressure gradient along the minor axis of 
the bubble, which yields a high pressure (fast expansion) at the major 
axis and a low pressure (slow expansion) at the minor axis. 
Magneto-hydrodynamical simulations (van der Swaluw, 2003; Komissarov and 
Lyubarsky, 2004; Del Zanna et al., 2004) have confirmed these results. 
Some of these simulations (Komissarov and Lyubarsky, 2004; Del Zanna et 
al., 2004) have been performed such that the condition of spherical symmetry 
is no longer enforced, instead an anisotropic pulsar wind, which is more
consistent with pulsar wind theory, has been used. These simulations yield
a more complex termination shock structure and post-shock flow. The X-ray 
synchrotron map calculated from these simulations by
Komissarov and Lyubarsky (2004) yields a similar jet-torus pattern as is 
observed in the X-ray Chandra map of the Crab Nebula.   

\section{Moving pulsars}

It is believed that pulsars can obtain a velocity at birth, due to for
example asymmetric supernova explosions. Recent studies (Arzoumanian et
al. 2002) show strong evidence for a two-component velocity distribution
of 90 and 500 km s$^{-1}$ for isolated radio pulsars. The velocity of a
pulsar will obviously influence the evolution of the PWN, as will be 
discussed below

In the supersonic expansion stage, the expansion velocity of the PWN
shock is high, i.e. typically 1,000-5,000 km/sec. These values
are much higher then typical values of pulsar velocities as mentioned
above. Therefore before the passage of the reverse shock, the evolution 
of the PWN will not be influenced by the motion of the pulsar.

The time for the reverse shock to collide with the complete PWN shock 
surface scales roughly linear with the velocity of the pulsar, as was 
shown by van der Swaluw et al. (2004). For large velocities, this
collision process can be a significant fraction of the total lifetime
of the remnant. The model from van der Swaluw et al. (2004) assume a 
constant uniform density of the ISM. In contrast, Blondin et al. (2001) 
performed simulations for which the pulsar is steady ($V_{\rm psr}=0$), 
whereas the ISM is non-uniform. They show that for such a configuration 
the collision of the reverse shock with the whole of the PWN shock surface
is also not instantaneous. It should be stressed that the exact collision 
process and its associated timescale for a PWN, will in general depend on a 
{\it combination} of the motion of the pulsar and the structure of  the 
surrounding medium in which the {\it SNR forward shock} propagates.
Both the simulations from Blondin et al. (2001) and van der Swaluw et al. 
(2004) show that after the crushing event of the PWN, there is a clear two-component
PWN structure:  the relic PWN, which contains a large fraction of the
injected electrons during the pulsar wind's lifetime, and the head of the
PWN, which is supplied by freshly injected electrons by the pulsar wind. 
The relic PWN would therefore radiate brightly at radio frequencies (also 
due to the increased magnetic field strength). The PWN head would be a weaker 
radio source but be bright at X-ray frequencies, which is lacking in the relic 
PWN, due to the short synchrotron lifetime of X-ray electrons.
N157B and G327.1-1.1 are two examples which clearly show such a
morphology, indicating that these two SNRs are examples of PWNe
for which the supersonic expansion stage has been terminated by
the passage of the reverse shock. 

The simulation from van der Swaluw et al. (2004) shows that after
the passage of the reverse shock, the head of the PWN, containing
the young pulsar will ultimately deform into a bow shock. This
transition occurs once the pulsar motion becomes larger compared
with the sound speed of the surrounding SNR material. Exact numbers
for this transition can be given when the pulsar propagates through 
a Sedov-Taylor remnant (van der Swaluw et al. 1998), i.e. the transition 
occurs when the position of the pulsar equals $\simeq 0.667$ times the 
blastwave radius of the SNR at roughly {\it half} the crossing time.
However, most pulsars are expected to take over their remnant's shell 
when the SNR has already made its transition to the pressure-driven 
snowplow stage, of which PSR1951+32 in the SNR CTB80 is an excellent 
example.

\begin{figure}
\includegraphics[width=140mm]{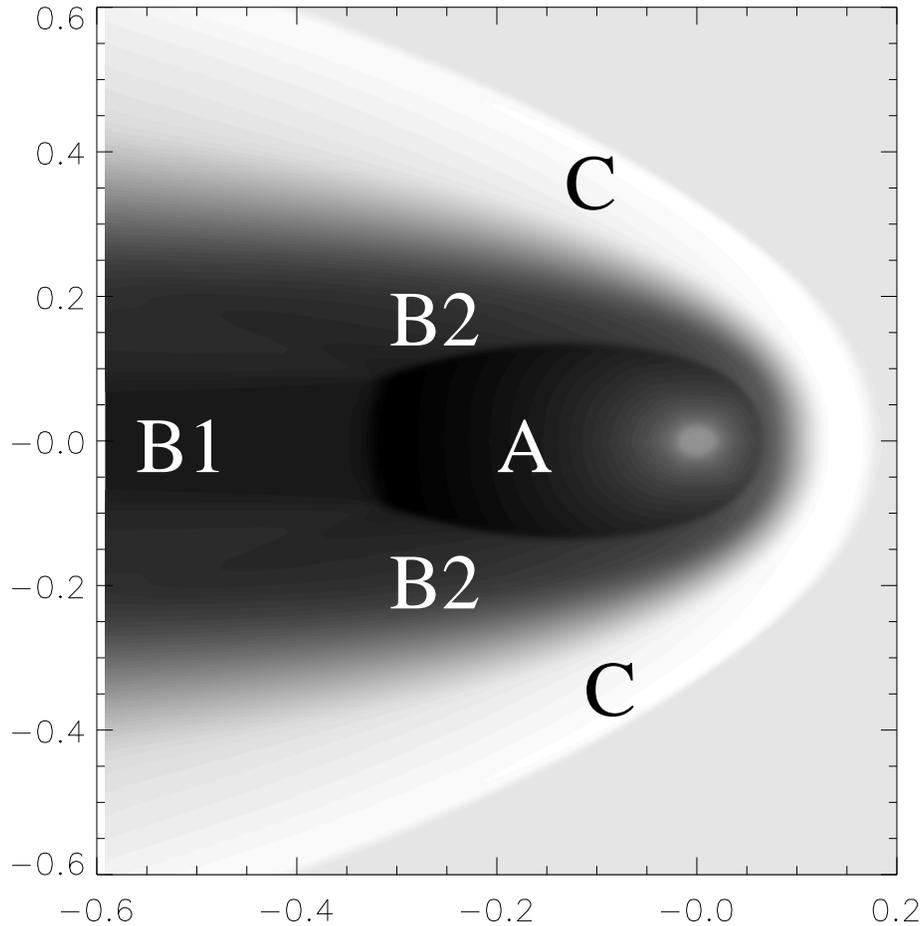}
\caption{Logarithmic density distribution of a pulsar wind nebula bow
shock. The simulation was performed to identify the different regions
(see text section 3) in the Mouse (Gaensler et al. 2004).}
\end{figure}

After the break-through event of a PWN bow shock through its associated 
SNR, the PWN is interacting directly with the ISM material.
Currently there are only
a few PWNe bow shocks known of which the Mouse (Gaensler et al.
2004) is the brightest one in both radio and X-rays. However, up till
now there has been no H$\alpha$ detection of this system. A nice
example of a bow shock nebula observed in both H$\alpha$ and X-rays
is the Black Widow around PSR B1957+20 (Stappers et al., 2003). The X-ray emission of these
type of bow shocks is believed to come from the shocked pulsar wind 
material, whereas the H$\alpha$ emission is originating from the
swept-up ISM material by the bow shock. Several hydrodynamical
simulations (Bucciantini 2002; van der Swaluw et al. 2003) have been
performed for these type of PWN bow shocks. Figure 4 depicts an
example of such a morphology of a PWN bow shock (from
Gaensler et al. 2004). The logarithmic 
density distribution is shown for a pulsar moving with a velocity
much higher than the ISM sound speed. The separate
regions in such a pulsar wind nebula bow shock can be distinguished:
A/ The pulsar wind cavity, which is elongated due to the supersonic
motion of the pulsar B/ The region containing the shocked pulsar
wind material, part of it is forming a tail (region B1 subsonic,
region B2 supersonic). C/ the swept-up ISM
material by the bow shock bounding the PWN bubble. In case of the
Mouse, these different areas have been identified with the X-ray map
from Chandra (Gaensler et al. 2004). 

\section{Prospects}

In recent years the amount of observational data on PWNe, due to
new observational facilities has increased enormously. At the same
time the rise of relativistic magneto-hydrodynamical simulations is
improving our theoretical understanding of PWNe. Future studies of
PWNe will hopefully lead to a comparison between multi-wavelength
studies and numerical models in order to explain the variety of
morphologies observed in plerionic SNRs.

\noindent
{\bf Acknowledgements.} I would like to thank the participants and
ISSI (International Space Science Institute, Bern) for a stimulating
workshop on the ``Physics of Supernova remnants in the XMM-Newton,
Chandra and INTEGRAL era'', which was helpful for writing this review
paper. 





\begin{thebibliography}{}


\item
Arzoumanian, Z., Chernoff, D. F., Cordes, J. M. The Velocity Distribution of 
Isolated Radio Pulsars. Astrophys. J. {\bf 568}, 289-301, 2002.

\item
Begelman, M. C., Li, Z.-Y, An axisymmetric magnetohydrodynamic model  for the 
Crab pulsar wind bubble, Astrophysical Journal, {\bf 397}, 187-195, 1992

\item
Blondin, J. M., R. A. Chevalier, and D. M. Frierson, Pulsar wind nebulae in evolved supernova remnants,
{\em Astrophys. J.}, {\bf 563}, 806-815, 2001.

\item
Bucciantini, N., Pulsar bow-shock nebulae. II. Hydrodynamical simulation,
Astronomy and Astrophysics, {\bf 387}, 1066-1073, 2002

\item
Bucciantini, N., Blondin, J. M., Del Zanna, L., Amato, E.,
Spherically symmetric relativistic MHD simulations of pulsar wind nebulae in supernova remnants,
Astronomy and Astrophysics, {\bf 405}, 617-626, 2003 

\item
Bucciantini, N., Bandiera, R., Blondin, J. M., Amato, E., Del Zanna, L.,
The effects of spin-down on the structure and evolution of pulsar wind nebulae,
Astronomy and Astrophysics, {\bf 422}, 609-619, 2004

\item
Chevalier, R. A., Fransson, C., Pulsar nebulae in supernovae,
Astrophysical Journal, {\bf 395}, 540-552, 1992

\item
Chevalier, R. A., Young core collapse supernova remnants and their
supernovae. The Astrophysical Journal, in press, astro-ph/0409013

\item
Cioffi, D. F., McKee, C. F., Bertschinger, E. Dynamics of radiative supernova 
remnants. The Astrophysical Journal 334, 252-265, 1988 

\item
Del Zanna, L., Amato, E., Bucciantini, N., Axially symmetric relativistic MHD simulations of 
Pulsar Wind Nebulae in Supernova Remnants. On the origin of torus and jet-like features,
Astronomy and Astrophysics, {\bf 421}, 1063-1073, 2004 

\item
 Gaensler, B. M., van der Swaluw, E., Camilo, F., Kaspi, V. M., Baganoff, F. K.,
 Yusef-Zadeh, F., Manchester, R. N. The Mouse That Soared: High Resolution 
X-ray Imaging of the Pulsar-Powered Bow Shock G359.23-0.82, Astrophys. J., in press
astro-ph/0312362

\item
Gallant, Y. A., van der Swaluw, E., Kirk, J. G., Achterberg, A. Modeling Plerion 
Spectra and their Evolution. in Neutron Stars in Supernova Remnants, ASP Conference 
Series, Vol. 271, 99-104, Eds P. O. Slane and B. M. Gaensler. San Francisco: ASP, 2002

\item
Garcia-Segura, G., Langer, N., Mac Low, M.-M. The hydrodynamic evolution of 
circumstellar gas around massive stars. II. The impact of the time sequence 
O star -$>$ RSG -$>$ WR star. Astronomy and Astrophysics 316, 133-146, 1996

\item
Heger, A., Fryer, C. L., Woosley, S. E., Langer, N., Hartmann, D. H. How 
Massive Single Stars End Their Life.The Astrophysical Journal 591, 288-300,
2003

\item
Hoshino, M., Arons, J., Gallant, Y. A., Langdon, A. B. Relativistic 
magnetosonic shock waves in synchrotron sources - Shock structure and 
nonthermal acceleration of positrons. Astrophys. J., {\bf 390}, 454-479,
1992

\item
Kaspi, V. M., Roberts, M. S. E., Harding, A. K. Isolated Neutron Stars.
in: Compact Stellar X-ray Sources", eds. W.H.G. Lewin and M. van der Klis,
{\it astro-ph/0402136} 

\item
Komissarov, S. S., Lyubarsky, Y. E., Synchrotron nebulae created by anisotropic
magnetized pulsar winds, {\bf 349}, 779-792, 2004

\item
McKee, C. F. X-Ray Emission from an Inward-Propagating Shock in Young Supernova 
Remnants. Astrophysical Journal, 188, 335-340, 1974

\item
McKee, C. F., Truelove, J. K. Explosions in the interstellar medium. Physics
Report 256, 157-172, 1995

\item
Kennel, C. F., Coroniti, F. V. Confinement of the Crab pulsar's wind by its supernova 
remnant, The Astrophysical Journal, {\bf 283}, 694-709, 1984

\item
Pacini, F., Salvati, M. On the Evolution of Supernova Remnants. Evolution of the 
Magnetic Field, Particles, Content, and Luminosity, Astrophysical Journal, {\bf 186}, 
249-266, 1973

\item
Rees, M. J., and J. E. Gunn, The origin of the magnetic field and relativistic particles 
in the Crab Nebula, {\em Mon. Not R. Astron. Soc.}, {\bf 167}, 1-12, 1974. 

\item
Reynolds, S. P., Chevalier, R. A. Evolution of pulsar-driven supernova remnants.
The Astrophysical Journal, 278, 630-648, 1984

\item
Stappers, B. W., Gaensler, B. M., Kaspi, V. M., van der Klis, M., Lewin, W. 
H. G., An X-ray nebula associated with the millisecond pulsar B1957+20.,
Science, 299, 1372-1374, 2003

\item
Truelove, J. K., McKee, C. F. Evolution of Nonradiative Supernova Remnants
The Astrophysical Journal Supplement 120, 299-326, 1999

\item
van der Swaluw, E., A. Achterberg, and Y. A. Gallant, 
Hydrodynamical simulations of pulsar wind nebulae in supernova remnants,
{\em Memorie della Societa Astronomia Italiana}, {\bf 69}, 1017-1022, 1998.

\item
van der Swaluw, E., A. Achterberg, Y. A. Gallant, and G. T\'oth,
Pulsar wind nebulae in supernova remnants. Spherically symmetric hydrodynamical simulations,
{\em Astron. Astrophys.}, {\bf 380}, 309-317, 2001.

\item
van der Swaluw, E., A. Achterberg, Y. A. Gallant, et al,
Interaction of high-velocity pulsars with supernova remnant shells, 
{\em Astron. Astrophys.}, {\bf 397}, 913-920, 2003

\item
van der Swaluw, E., Interaction of a magnetized pulsar wind with its surroundings. 
MHD simulations of pulsar wind nebulae, Astronomy and Astrophysics, {\bf 404}, 
939-947, 2003

\item
van der Swaluw, E., Downes, T. P., Keegan, R. An evolutionary model for pulsar-driven 
supernova remnants. A hydrodynamical model, Astronomy and Astrophysics, {\bf 420}, 
937-944, 2004

\end{thebibliography}
\end{document}